\documentclass[graybox]{svmult}

\usepackage{mathptmx}       
\usepackage{helvet}         
\usepackage{courier}        
\usepackage{type1cm}        
\usepackage{makeidx}         
\usepackage{graphicx}        
\usepackage{multicol}        
\usepackage[bottom]{footmisc}

\makeindex             

\begin{document}

\title*{Tidal Stirring of Milky Way Satellites:\\ A Simple Picture with the Integrated Tidal Force}
\titlerunning{Tidal Stirring of Milky Way Satellites}
\author{Ewa L. {\L}okas, Stelios Kazantzidis, Lucio Mayer and Simone Callegari}
\authorrunning{E. L. {\L}okas et al.}
\institute{
Ewa L. {\L}okas \at Nicolaus Copernicus Astronomical Center, 00-716 Warsaw, Poland, \email{lokas@camk.edu.pl}
\and
Stelios Kazantzidis \at Center for Cosmology and Astro-Particle Physics; and Department of Physics;
and Department of Astronomy, The Ohio State University, Columbus, OH 43210, USA, \email{stelios@mps.ohio-state.edu}
\and
Lucio Mayer \at Institute for Theoretical Physics, University of Z\"urich, CH-8057 Z\"urich, Switzerland,
\email{lucio@phys.ethz.ch}
\and
Simone Callegari \at Institute for Theoretical Physics, University of Z\"urich, CH-8057 Z\"urich, Switzerland,
\email{callegar@physik.uzh.ch}
}
\maketitle

\vskip-1.2truein

\abstract{Most of dwarf spheroidal galaxies in the Local Group were probably formed via environmental processes
like the tidal interaction with the Milky Way. We study this process via $N$-body simulations of dwarf galaxies
evolving on seven different orbits around the Galaxy. The dwarf galaxy is initially
composed of a rotating stellar disk and a dark matter halo. Due to the action of tidal forces it loses mass
and the disk gradually transforms into a spheroid while stellar motions become increasingly random.
We measure the characteristic scale-length
of the dwarf, its maximum circular velocity, mass, shape and kinematics as a function of the integrated tidal
force along the orbit. The final properties of the evolved dwarfs are remarkably similar if the total tidal
force they experienced was the same, independently of the actual size and eccentricity of the orbit.}

\section{Introduction}

In the tidal stirring scenario for the formation of dwarf spheroidal (dSph) galaxies \cite{may01} the progenitors
are late type dwarfs affected by tidal forces from the Milky Way or any other normal size galaxy. A similar process
may lead to the formation of S0s in galaxy clusters \cite{gne03a, gne03b}.
In the absence of any analytical description of tidal effects for dwarfs on eccentric orbits, that seem to dominate
in $\Lambda$CDM cosmologies, the problem can be fully addressed only via $N$-body simulations. With this tool, the
tidal stirring scenario has been recently tested for a variety of orbits and structural parameters of the dwarfs
and shown to work very efficiently towards the formation of dSph galaxies \cite{may07, kli09, lok10a, kaz10, lok10b}.
It was demonstrated that during the tidal evolution the dwarfs lose mass via tidal stripping, they undergo
morphological transformation from disks to spheroids and the rotation of their stars is replaced by random motions.

The exact mechanism behind this transformation still eludes us however. It is generally believed that the processes
shaping the dwarf galaxies orbiting on eccentric orbits may be inherently different from those on circular orbits due
to the time-dependence of the tidal force. While on circular orbits tidal forces are believed to mainly steepen
the outer density profile, on eccentric orbits they likely induce strong shocks to the whole structure of the dwarf
galaxy at pericenters \cite{bt08}. It has also been suggested however \cite{may01} that the key factor that controls
the extent of transformation is the integrated tidal force the dwarf experiences along the orbit rather than
the particular shape of the orbit. Here we address this question using a subset of collisionless $N$-body simulations
described in detail in \cite{kaz10}.

\begin{table}[t]
\caption{Orbital parameters of the simulated dwarfs}
\label{parameters}
\begin{tabular}{p{2cm}p{2cm}p{2cm}p{2cm}p{2cm}p{1cm}}
\hline\noalign{\smallskip}
Orbit & $r_{\rm apo}$ [kpc] & $r_{\rm peri}$ [kpc] & $T_{\rm orb}$ [Gyr] & $t_{\rm la}$ [Gyr] & $n_{\rm apo}$ \\
\noalign{\smallskip}\svhline\noalign{\smallskip}
O1  &    125  &  25  &  2.09  & \ \  8.35  &  5 \\
O2  & \ \ 87  &  17  &  1.28  & \ \  8.95  &  8 \\
O3  &    250  &  50  &  5.40  & \ \  5.40  &  2 \\
O4  &    125  & 12.5 &  1.81  & \ \  9.05  &  6 \\
O5  &    125  &  50  &  2.50  &     10.00  &  5 \\
O6  & \ \ 80  &  50  &  1.70  & \ \  8.50  &  6 \\
O7  &    250  & 12.5 &  4.55  & \ \  9.10  &  3 \\
\noalign{\smallskip}\hline\noalign{\smallskip}
\end{tabular}
\end{table}

In these numerical experiments a dwarf galaxy, initially composed of a stellar disk and a dark matter halo, is placed on
seven different orbits O1-O7 around a live Milky Way model. The orbital parameters of the simulations are listed in
Table~\ref{parameters}. Orbits O1-O5 correspond to runs R1-R5 in \cite{kaz10}, while orbits O6 and O7 are two additional
setups. The second and third column of the Table list the apo- and pericenter distances and the fourth one
the orbital time.
All simulations were evolved for 10 Gyr; the fifth column gives the time when the last apocenter occurred and the
last column the total number of apocenters.

Our dwarf galaxy model was identical in all runs (model D1 in \cite{kaz10}). It contained a stellar disk of mass
$M_{\rm d} = 2 \times 10^7$ M$_{\odot}$ with scale-length $R_{\rm d} = 0.41$ kpc and thickness parameter
$R_{\rm d}/z_{\rm d} = 0.2$. The dwarf's dark matter halo had an NFW profile with virial mass
$M_{\rm h} = 10^9$ M$_{\odot}$ and concentration $c=20$. The mass distribution of the Milky Way was given
by model MWb in \cite{wd05}.

\section{Results}

\begin{figure}[t]
\includegraphics[scale=0.78]{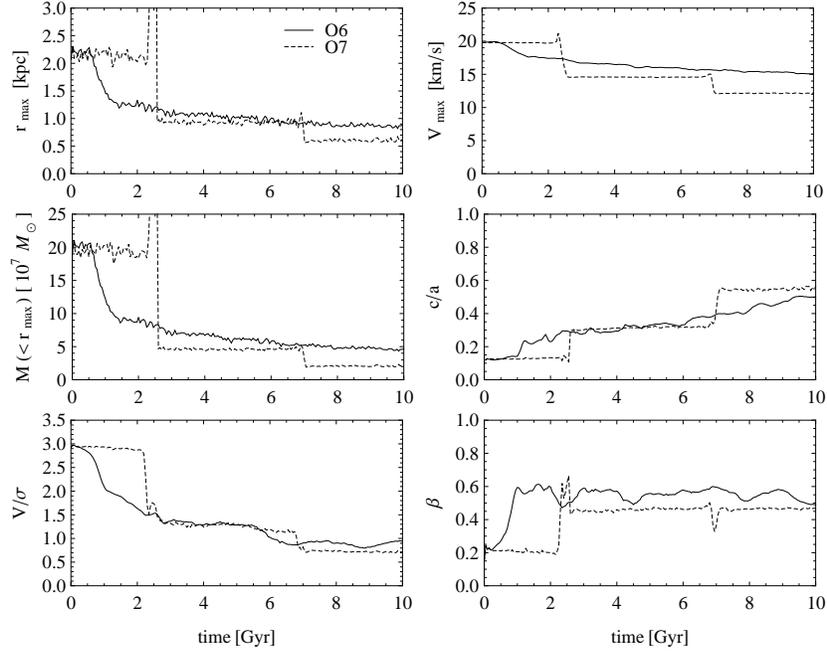}
\caption{Evolution of the properties of the dwarf as a function of time for the least (O6, solid line)
and the most (O7, dashed line) eccentric orbit.}
\label{lokas_fig1}
\end{figure}

For each of the simulations we calculated the evolution of a few key properties of the dwarf in time: the characteristic
radius $r_{\rm max}$ at which the circular velocity has a maximum, the maximum circular velocity $V_{\rm max}$,
and the mass contained within the characteristic scale, $M(<r_{\rm max})$. Using stars inside $r<r_{\rm max}$ we
determined the shape and kinematic properties of the stellar component. The shape is described in terms of the
shortest to longest axis ratio $c/a$ calculated from the
moments of the inertia tensor. The amount of ordered versus random motion was quantified by the ratio of the
rotation velocity around the shortest axis to the 1D velocity dispersion, $V/\sigma$. The anisotropy of the
stellar motions was characterized by the usual $\beta$ parameter. The evolution of these properties as
a function of time is illustrated in Figure~\ref{lokas_fig1} for the least and the most eccentric orbit (O6 and O7).
The results for other orbits are discussed in \cite{kaz10}.

\begin{figure}
\includegraphics[scale=0.85]{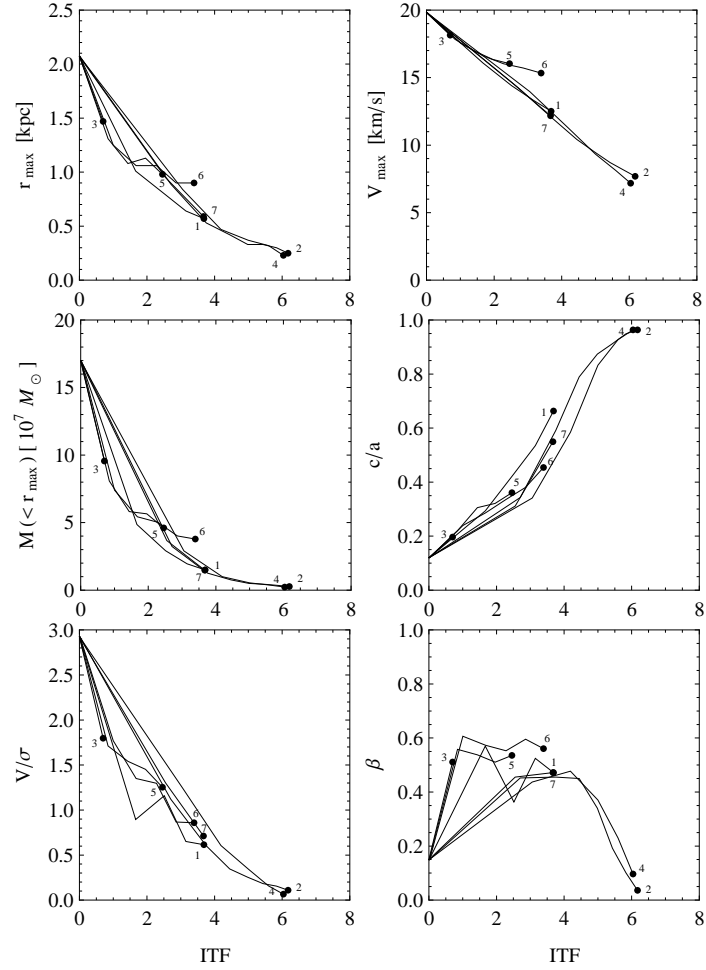}
\caption{Evolution of the properties of the dwarf as a function of the integrated tidal force (ITF, in arbitrary
units). The lines connect measurements at subsequent apocenters for a given orbit with dots marking the results
at the last apocenter. In each panel all lines start from the same point at ITF=0 because the initial structural
properties of the dwarf were the same for all orbits and all simulations started at apocenters.}
\label{lokas_fig2}
\end{figure}

Next, we calculated the tidal force experienced by the dwarf galaxy on different orbits as
\begin{equation}      \label{tidal force}
	F_{\rm tidal} \propto r_{\rm max} M(<R)/R^3\;,
\end{equation}
where $r_{\rm max}$ is the characteristic scale-length of the dwarf at which the tidal force operates,
$R$ is the distance between the center
of the dwarf and the center of the Milky Way and $M(<R)$ is the mass of the Milky Way model within this distance
(approximated as a spherical NFW distribution with the disk and the bulge added as point masses).
Summing up contributions of this form over a given orbit we obtain an estimate of the integrated tidal force
(ITF) experienced by the dwarf up to a given time. To make the results comparable for different orbits it is
advisable to integrate over full orbital times so for each of the orbits we summed up the tidal force from the first
up to the last apocenter.

The evolution of different properties of the dwarf, described above, as a function of the integrated tidal force,
is shown in Figure~\ref{lokas_fig1}. The lines join the results for a given orbit at subsequent apocenters with
the dots marking the last apocenter. Numbers near the points indicate the orbit, with 1-7 corresponding to
orbits O1-O7 in Table~\ref{parameters}. The values of the parameters of the dwarf at the last apocenter are listed
in Table~\ref{properties}.

\section{Conclusions}

The tracks of the dwarf galaxy properties as a function of the integrated tidal force are remarkably similar.
In spite of different size and eccentricity of the orbits the evolution seems to be controlled
almost entirely by the amount of tidal force experienced by the dwarf. The similarity is particularly striking
in the case of two pairs of orbits, O2-O4 and O1-O7. Although the orbital parameters in each pair are very different
(see Table~\ref{parameters}) the dwarfs experience a very similar integrated tidal force and all their properties at
the last apocenter are also almost identical. A slight departure from the trend set by the orbits O2-O4 and O1-O7
is seen only in the case of $V_{\rm max}$ for orbits O5-O6 with the largest pericenter.

\begin{table}[t]
\caption{Properties of the simulated dwarfs at the last apocenter}
\label{properties}
\begin{tabular}{p{1cm}p{2cm}p{2cm}p{3cm}p{1cm}p{1cm}p{1cm}}
\hline\noalign{\smallskip}
Orbit & $r_{\rm max}$ [kpc] & $V_{\rm max}$ [km/s] & $M(<r_{\rm max})$ [$10^7$ M$_{\odot}$]
& $c/a$ & $V/\sigma$ & $\beta$ \\
\noalign{\smallskip}\svhline\noalign{\smallskip}
O1  & 0.57    &    12.5     &  1.50   &  0.66   &  0.62   &   0.47    \\
O2  & 0.25    & \ \ 7.7     &  0.27   &  0.96   &  0.11   &   0.04    \\
O3  & 1.47    &    18.1     &  9.56   &  0.20   &  1.80   &   0.51    \\
O4  & 0.23    & \ \ 7.2     &  0.22   &  0.96   &  0.07   &   0.10    \\
O5  & 0.98    &    16.0     &  4.61   &  0.36   &  1.25   &   0.54    \\
O6  & 0.90    &    15.3     &  3.79   &  0.45   &  0.86   &   0.56    \\
O7  & 0.59    &    12.2     &  1.48   &  0.55   &  0.71   &   0.47    \\
\noalign{\smallskip}\hline\noalign{\smallskip}
\end{tabular}
\end{table}

\begin{acknowledgement}
This research and EL{\L}'s participation in JENAM 2010 was partially
supported by the Polish Ministry of Science and Higher Education under grant NN203025333.
\end{acknowledgement}


\begin{thebibliography}{99.}

\bibitem{bt08} Binney J., Tremaine S., 2008, Galactic Dynamics. Princeton Univ. Press, Princeton

\bibitem{gne03a} Gnedin O. Y., 2003, ApJ, 582, 141

\bibitem{gne03b} Gnedin O. Y., 2003, ApJ, 589, 752

\bibitem{kaz10} Kazantzidis S., {\L}okas E. L., Callegari S., Mayer L., Moustakas L.
        A., 2010, ApJ, in press, \url{http://arxiv.org/abs/1009.2499}

\bibitem{kli09} Klimentowski J., {\L}okas E. L., Kazantzidis S., Mayer L., Mamon G. A.,
	2009, MNRAS, 397, 2015

\bibitem{lok10a} {\L}okas E. L., Kazantzidis S., Klimentowski J., Mayer L., Callegari S.,
        2010, ApJ, 708, 1032

\bibitem{lok10b} {\L}okas E. L., Kazantzidis S., Majewski S. R., Law D. R., Mayer L., Frinchaboy P. M.,
	2010, ApJ, in press, \url{http://arxiv.org/abs/1008.3464}

\bibitem{may01} Mayer L., Governato F., Colpi M., Moore B., Quinn T., Wadsley J.,
        Stadel J., Lake G., 2001, ApJ, 559, 754

\bibitem{may07} Mayer L., Kazantzidis S., Mastropietro C., Wadsley J., 2007,
        Nature, 445, 738

\bibitem{wd05} Widrow L. M., Dubinski J., 2005, ApJ, 631, 838


\end{thebibliography}
\end{document}